\documentclass[12pt,eqsecnum]{revtex4}
\begin{document}

\title{Generalized BRST transformation in Abelian rank-2 antisymmetric tensor field theory}

\author{ Sudhaker Upadhyay\footnote {e-mail address: sudhakerupadhyay@gmail.com}}
\author{ Bhabani Prasad Mandal\footnote{e-mail address:
bhabani.mandal@gmail.com,\ \ bhabani@bhu.ac.in  }}

\affiliation { Department of Physics,\\
Banaras Hindu University,\\
Varanasi-221005, INDIA. \\
}

\begin{abstract}
We generalize the BRST transformations in Abelian rank-2 tensor field theory by allowing the 
parameter to be finite and field dependent and show that such transformations play  crucial 
role in studying the Abelian 2-form gauge theory in noncovariant 
gauges. The generating functionals in different effective theories are connected through 
these generalized BRST transformations with different choice of finite parameter. We further 
consider the generalization of the anti-BRST transformations 
 and show that even anti-BRST transformations with the finite field dependent parameter 
relate the generating functional for different effective theories. Such generalized 
BRST/anti-BRST transformations are useful in connecting generating functionals corresponding 
to different effective theories. Field/antifield formulation of this Abelian 2-form gauge 
theory is also studied in the context of generalized BRST transformations. 
We show that generalized BRST transformation connects the generating functional
corresponding to the different solutions of quantum master equation in field/antifield 
formulation.
\end{abstract}
\maketitle

{\it keywords}: {2-form gauge theory; BRST transformation; Field/Antifield formulation.}

\section{Introduction}	

Gauge theories of Abelian rank-2 antisymmetric tensor field play crucial role in studying the 
theory for classical strings \cite{a}, vortex motion in an irrotational, incompressible 
fluid \cite{b,c} and the dual formulation of the Abelian Higgs model \cite{d,e}. 
Abelian 
rank-2 antisymmetric tensor fields are also very useful in studying supergravity multiplets 
\cite{g}, excited states in superstring theories \cite{h,i} and anomaly cancellation in 
certain superstring theories. Geometrical aspects of Abelian rank-2 antisymmetric tensor 
fields are studied in a U(1) gauge theory in loop space. The Ward-Takahashi identities  are 
derived for such theories in a superspace formulation \cite{f}.

Covariant quantization using BRST formulation for an Abelian rank-2 antisymmetric tensor 
field was studied consistently by many people \cite{j,k,l,m,n,o}. In such covariant 
quantization a naive gauge fixing term containing the antisymmetric tensor field is itself 
invariant under a secondary gauge transformation and hence further commuting ghost fields 
(ghost of ghosts) are required for complete gauge fixing. The usual BRST transformations are 
characterized by an infinitesimal, anticommuting and global parameter. These 
transformations in pure Yang-Mills theory have been generalized by allowing the parameter to 
be finite, field dependent
 but independent of $x_\mu$ explicitly (known as FFBRST transformations)\cite{sdj}. Such a 
generalized BRST transformations are symmetry of the effective action and are nilpotent.

 The generalized BRST transformations have found many applications in studying 1-form gauge 
theory \cite{sdj,sdj01,sdj02,sdj4,sdj03,sdj04,sdj05,rb,sdj5,sm,ssm}. A correct and explicit 
prescription for 
the poles in the gauge field propagators in noncovariant gauges have been derived by 
connecting the effective theories in covariant gauge to the theories in noncovariant gauge 
through FFBRST transformations \cite{sdj01,sdj5}. Hence one does not need to put a 
prescription for the poles by hand in the formulation using FFBRST transformations 
\cite{sdj01}.
 The divergent energy integrals in Coulomb 
gauge are regularized by modifying the time like propagator by considering 
the generalized BRST 
transformations \cite{sdj5}. The question of gauge invariance of the Wilson loop
in the light of a new treatment of axial gauge propagator \cite{sdj01} are proposed
 based on a the FFBRST transformations \cite{sdj03,sdj04,sdj05}. Using the technique 
of generalized BRST transformations it has been shown that the classical massive Yang-Mills 
theory and the pure Yang-Mills
theory whose gauge symmetry is broken by a gauge fixing term are identical
from the view point of quantum gauge symmetry \cite {rb}. 

In this present work we study the quantization of the Abelian rank-2 antisymmetric tensor 
field by using the generalized BRST transformations. We have shown that it is possible to 
construct the Abelian rank-2 tensor field theory in noncovariant gauges by using generalized 
BRST transformations. In particular, we have shown the generating functional for Abelian 
rank-2 tensor field theory in covariant gauges transformed to the generating functional for 
the same theory in a noncovariant gauges for a particular choice of the finite parameter in 
generalized BRST transformations. The new results arise from the nontrivial Jacobian of the 
path integral measure under such finite BRST transformations. We have established the 
connections explicitly for the theories in two different noncovariant gauge, namely axial 
gauge and 
Coulomb gauge.

The Lagrangian quantization of Batalin and Vilkovisky \cite{bv,bv1,bv2,ht,wei}, also known as 
the 
field/antifield formulation considered to be one of the most powerful and advanced technique
of quantization of gauge theories involving the BRST symmetry \cite{ht,wei}. We consider 
the field/antifield formulation of Abelian rank-2 tensor field theory by introducing 
the antifield $\phi^\star $ corresponding to each field $\phi$ with opposite statistics
to study the role of generalized BRST transformation in such formulation. We show that 
the FFBRST
transformation 
changes the generating functional corresponding to one gauge fixed fermion to the generating 
functional to another gauge fixed fermion. Thus FFBRST transformation connects the different 
solutions of 
the master equation in field/antifield formulation. We show this by considering an explicit 
example.

This paper is organized in the following manner. We start with a brief discussion about 
Abelian rank-2 antisymmetric tensor field theory in the Sec. 2. A brief introduction of the 
generalized BRST transformations is given in Sec. 3. The connections between different 
generating functionals
through generalized BRST transformations in 2-form gauge theories
with appropriate examples are shown in the Sec. 4. 
The generalized anti-BRST transformations 
for such theories are considered in the Sec. 5. The field/antifield formulation of 
2-form gauge theory using FFBRST transformation
is mentioned in 
Sec. 6. Mathematical details, for section 4, are 
provided in Appendices. Last section is kept for discussion and concluding remarks.

\section{\bf Preliminary : Gauge theory of Abelian rank-2 Antisymmetric Tensor field}

We consider the Abelian gauge theory for rank-2 antisymmetric tensor field $B_{\mu\nu}$ 
defined by the action
\begin{equation}
S_0=\frac{1}{12}\int d^4x F_{\mu \nu \rho}F^{\mu \nu \rho},\label{kin}
\end{equation}
where $F_{\mu \nu \rho}\equiv \partial_\mu B_{\nu\rho}+\partial_\nu B_{\rho\mu}+\partial_\rho 
B_{\mu\nu}.$ This action is invariant under the gauge transformation $\delta 
B_{\mu\nu}=\partial_{\mu}\zeta_{\nu} -\partial_{\nu}\zeta_{\mu}$ with a vector gauge 
parameter $\zeta_{\mu}(x)$.

To quantize this theory using BRST transformation, it is necessary to introduce the following 
ghost and auxiliary fields: anticommuting vector fields $\rho_{\mu}$ and $\tilde\rho_{\mu}$, 
a commuting vector field $\beta_{\mu}$, anticommuting scalar fields $\chi$ and $\tilde\chi$, 
and commuting scalar fields $\sigma, \varphi,$ and $ \tilde\sigma $. The BRST transformation 
is then defined for $B_{\mu\nu}$ by replacing $\zeta_{\mu}$ in the gauge transformation by 
the ghost field $\rho_{\mu}$.

The complete effective action for this theory in covariant gauge, using  the BRST 
formulation, is given by
\begin{equation}
S^L_{eff}=S_0+S_{gf}+S_{gh} \label{act}
\end{equation}
with the gauge fixing and ghost term
\begin{eqnarray}
S_{gf}+S_{gh}&=&\int d^4x\left[-i\partial_\mu\tilde\rho_\nu (\partial^\mu\rho^\nu -
\partial^\nu\rho^\mu )+\partial_\mu\tilde\sigma\partial^\mu\sigma +\beta_\nu(\partial_\mu B^{
\mu\nu} +\lambda_1\beta^\nu -\partial^\nu\varphi)\right.\nonumber\\ 
&-&\left. i\tilde\chi\partial_\mu\rho^\mu -i\chi (\partial_\mu\tilde\rho^\mu -
\lambda_2\tilde\chi)\right], \label{gfix}
\end{eqnarray}
where $\lambda_1$ and $\lambda_2$ are gauge parameters.
This effective action is invariant under following BRST and anti-BRST symmetries.

BRST:
\begin{eqnarray}
\delta_b B_{\mu\nu} &=& -(\partial_\mu\rho_\nu -\partial_\nu\rho_\mu)\Lambda \nonumber\\
\delta_b\rho_\mu &=& -i\partial_\mu\sigma \Lambda,  \ \ \ \ \ \ \ \ \ \ \ \ \ \ \ \delta_b\sigma 
= 0 \nonumber\\
\delta_b\tilde\rho_\mu &=&i\beta_\mu \Lambda,   \ \ \ \ \ \ \ \ \ \ \ \ \ \ \ \ \ 
\delta_b\beta_\mu = 0\nonumber\\
\delta_b\tilde\sigma &=&-\tilde\chi\Lambda,  \ \ \ \ \ \ \ \ \ \ \ \ \ \ \ \ \ \ \ 
\delta_b\tilde\chi =0\nonumber\\
\delta_b\varphi &=& -\chi\Lambda, \ \ \ \ \ \ \ \ \ \ \ \ \ \ \ \ \ \ \ \delta_b\chi =0,\label{sym}
\end{eqnarray}
Anti-BRST:
\begin{eqnarray}
\delta_{ab} B_{\mu\nu}&=&-(\partial_\mu\tilde\rho_\nu -\partial_\nu\tilde\rho_\mu)\Lambda 
\nonumber\\
\delta_{ab}\tilde\rho_\mu &=& -i\partial_\mu\tilde\sigma \Lambda,\ \ \ \ \ \ \ \ \ \ \ \ \ \ \ \ 
\delta_{ab}\tilde\sigma = 0 \nonumber\\
\delta_{ab}\rho_\mu &=&-i\beta_\mu \Lambda,\ \ \ \ \ \ \ \ \ \ \ \ \ \ \ \ \delta_{ab}\beta_\mu = 0
\nonumber\\
\delta_{ab}\sigma &=&-\chi\Lambda,\ \ \ \ \ \ \ \ \ \ \ \ \ \ \ \ \ \ \ \ \delta_{ab}\chi =0
\nonumber\\
\delta_{ab}\varphi &=& \tilde\chi\Lambda,\ \ \ \ \ \ \ \ \ \ \ \ \ \ \ \ \ \ \ \ \ \ 
\delta_{ab}\tilde\chi =0,\label{asym}
\end{eqnarray}
where the BRST parameter $\Lambda$ is global, infinitesimal and anticommuting in nature.
The anti-BRST transformations are similar to the BRST transformations, where the role of 
ghost and antighost field is interchanged with some modification in coefficients.
The generating functional for this Abelian rank 2 antisymmetric tensor field theory in a
covariant gauge is defined as,
\begin{equation}
Z^L=\int D\phi\exp [iS_{eff}^L[\phi ]]\label{zfun}
\end{equation}
where $\phi$ is the generic notation for all the fields ($B_{\mu\nu}, \rho_\mu, 
\tilde\rho_\mu, \beta_\mu, \varphi, \sigma, \tilde\sigma, \chi, \tilde\chi$).

The BRST and anti-BRST transformations in Eqs. (\ref{sym}) and (\ref{asym}) respectively 
leave the
above generating functional invariant as, the path integral measure $D\phi \equiv DB D\rho 
D\tilde\rho D\beta D\sigma D\varphi D\tilde\sigma D\chi D\tilde\chi$ is invariant under such 
transformations.
\section{Generalized BRST in Abelian rank 2 Anti-symmetric Tensor field}
The properties of the usual BRST transformations in Eq. (\ref{sym})  do not depend on whether 
the parameter $\Lambda$  is (i) finite or infinitesimal, (ii) field dependent or not, as long 
as it is anticommuting and space-time independent. These observations give us a freedom to 
generalize the BRST transformations by making the parameter, $\Lambda$ finite and field 
dependent without affecting its properties. To generalized the BRST transformations we start 
with making the  infinitesimal parameter field dependent by introducing a parameter $\kappa\ 
(0\leq \kappa\leq 1)$ and making all the fields, $\phi(x,\kappa)$, $\kappa$ dependent such that $\phi(x,\kappa =0)=\phi(x)$ and $\phi(x,\kappa 
=1)=\phi^\prime(x)$, the transformed field.

The usual infinitesimal BRST transformations, thus can be written generically as 
\begin{equation}
{d\phi(x,\kappa)}=\delta_{b} [\phi (x,\kappa ) ]\Theta^\prime [\phi (x,\kappa ) ]{d\kappa}
\label{diff}
\end{equation}
where the $\Theta^\prime [\phi (x,\kappa ) ]{d\kappa}$ is the infinitesimal but field 
dependent parameter.
The generalized BRST transformations with the finite field dependent parameter then can be 
constructed by integrating such infinitesimal transformations from $\kappa =0$ to $\kappa= 1$
, to obtain
\begin{equation}
\phi^\prime\equiv \phi (x,\kappa =1)=\phi(x,\kappa=0)+\delta_b[\phi(x) ]\Theta[\phi(x) ]
\label{kdep}
\end{equation}
where 
\begin{equation}
\Theta[\phi(x)]=\int_0^1 d\kappa^\prime\Theta^\prime [\phi(x,\kappa^\prime)],
\end{equation}
 is the finite field dependent parameter. 

It can easily be shown that such offshell nilpotent BRST transformations with finite field 
dependent parameter are symmetry  of the effective action in Eq. (\ref{act}). However, the 
path integral measure in Eq. (\ref{zfun}) is not invariant under such transformations as the 
BRST parameter is finite.

The Jacobian of the path integral measure for such transformations can be evaluated for some 
particular choices of the finite field dependent parameter, $\Theta[\phi(x)]$, as
\begin{eqnarray}
DB^\prime D\rho^\prime D\tilde\rho^\prime D\beta^\prime D\sigma^\prime D\tilde\sigma^\prime D
\chi^\prime D\tilde\chi^\prime &=&J(
\kappa) DB(\kappa) D\rho(\kappa) D\tilde\rho(\kappa) D\beta(\kappa) D\sigma(\kappa) D
\tilde\sigma(\kappa )\nonumber\\
&&D\chi(\kappa) D\tilde\chi(\kappa).
\end{eqnarray}
The Jacobian, $J(\kappa )$ can be replaced (within the functional integral) as
\begin{equation}
J(\kappa )\rightarrow \exp[iS_1[\phi(x,\kappa) ]]
\end{equation}
 iff the following condition is satisfied \cite{sdj}
\begin{equation}
\int {\cal{D}}\phi (x) \;  \left [ \frac{1}{J}\frac{dJ}{d\kappa}-i\frac
{dS_1[\varphi (x,\kappa )]}{d\kappa}\right ]\exp{[i(S_{eff}+S_1)]}=0 \label{mcond}
\end{equation}
where $ S_1[\phi ]$ is local functional of fields.

The infinitesimal change in the $J(\kappa)$ can be written as
\begin{equation}
\frac{1}{J}\frac{dJ}{d\kappa}=-\int d^4x\left [\pm \delta_b \phi (x,\kappa )\frac{
\partial\Theta^\prime [\phi (x,\kappa )]}{\partial\phi (x,\kappa )}\right ],\label{jac}
\end{equation}
where $\pm$ sign refers to whether $\phi$ is a bosonic or a fermionic field.

By choosing appropriate $\Theta$, we can make $S_{eff}+S_1$ either another effective action 
for same theory or effective action for another theory. The resulting effective action also 
be invariant under same BRST/anti-BRST transformations.
\section{ Generalized BRST in 2-form gauge theory : examples}

In this section we would like to show explicitly that the generalized BRST transformations 
can relate the different effective theories. In particular we are interested to obtain the 
effective theories for Abelain rank-2 tensor field in noncovariant gauges by applying 
generalized BRST transformations to the effective theories in covariant gauge.
\subsection{Effective theory in axial gauge}
We start with the generating functional corresponding to the effective theory in Lorentz gauge
given in Eq. (\ref{zfun}), where
$S^L_{eff}$ is invariant under following generalized BRST:
\begin{eqnarray} 
\delta_b B_{\mu\nu} &=& -(\partial_\mu\rho_\nu -\partial_\nu\rho_\mu)\Theta_b [\phi ] \nonumber\\
\delta_b\rho_\mu &=& -i\partial_\mu\sigma \Theta_b [\phi ],  \ \ \ \ \ \ \ \ \ \ \ \ \ \ \ 
\delta_b\sigma = 0 \nonumber\\
\delta_b\tilde\rho_\mu &=&i\beta_\mu \Theta_b [\phi ],   \ \ \ \ \ \ \ \ \ \ \ \ \ \ \ \ \ \ 
\delta_b\beta_\mu = 0\nonumber\\
\delta_b\tilde\sigma &=&-\tilde\chi\Theta_b [\phi ],  \ \ \ \ \ \ \ \ \ \ \ \ \ \ \ \ \ \ \ \ 
\delta_b\tilde\chi =0\nonumber\\
\delta_b\varphi &=& -\chi\Theta_b [\phi ], \ \ \ \ \ \ \ \ \ \ \ \ \ \ \ \ \ \ \ \ \delta_b\chi 
=0,\label{fsym}
\end{eqnarray}
where $\Theta_b$ is a finite, anticommuting BRST parameter depends on the fields in global manner.
To obtain the effective theories in axial gauge we choose the finite parameter as 
\begin{eqnarray}
\Theta^\prime_b &=&\int d^4x\left[\gamma_1\tilde\rho_\nu (\partial_\mu B^{\mu\nu}-\eta_\mu B^{
\mu\nu}-\partial^\nu\varphi -\eta^\nu\varphi )+\gamma_2 \lambda_1\tilde\rho_\nu\beta^\nu 
\right. \nonumber\\
&+&\left.\gamma_1\tilde\sigma (\partial_\mu\rho^\mu -\eta_\mu\rho^\mu )+\gamma_2 
\lambda_2\tilde\sigma\chi\right]\label{afin}
\end{eqnarray}
where $\gamma_1$ and $\gamma_2$ are arbitrary  parameters (depend on $\kappa$) and 
$\eta_\mu$ is arbitrary
constant four vector.
Now we apply these generalized BRST transformations to the generating functional $Z^L$ given in 
Eq. (\ref{zfun}). The path integral measure is not invariant  and  give rise a nontrivial 
functional $e^{iS_1^A}$ (explicitly shown in Appendix A), 
where 
\begin{eqnarray}
S_1^A&=&\int d^4x [-\beta_\nu\partial_\mu B^{\mu\nu} +\beta_\nu\eta_\mu B^{\mu\nu} +\gamma_2
 \lambda_1\beta_\nu\beta^\nu -i\tilde\rho_\nu\partial_{\mu}(\partial^\mu\rho^\nu -
\partial^\nu\rho^\mu)\nonumber\\
&+&i\tilde\rho_\nu\eta_{\mu}(\partial^\mu\rho^\nu -\partial^\nu\rho^\mu )+i\gamma_2 
\lambda_2\chi\tilde\chi +i\tilde\chi\partial_\mu\rho^\mu -i
\tilde\chi\eta_\mu\rho^\mu\nonumber\\
&+&\tilde\sigma\partial_\mu\partial^\mu\sigma -\tilde\sigma\eta_\mu\partial^\mu\sigma - 
\partial_\mu\beta^\mu\varphi + \eta_\mu\beta^\mu\varphi +i\chi\partial_\mu\tilde\rho^\mu 
\nonumber\\
&-&i\chi\eta_\mu\tilde\rho^\mu ].\label{s1}
\end{eqnarray}
Now, adding this $S_1^A$ to $S^L_{eff}$, we get 
\begin{equation}
S^L_{eff} +S_1^A\equiv S^A_{eff},
\end{equation}
where
\begin{eqnarray}
S^A_{eff}&=&\int d^4x\left[\frac{1}{12}F_{\mu\nu\rho}F^{\mu\nu\rho}+i\tilde\rho_\nu\eta_\mu (
\partial^\mu\rho^\nu -\partial^\nu\rho^\mu )-\tilde\sigma\eta_\mu\partial^\mu\sigma +
\beta_\nu(\eta_\mu B^{\mu\nu}\right.\nonumber\\ 
&+&\left. \lambda_1^\prime\beta^\nu +\eta^\nu\varphi) 
- i\tilde\chi\eta_\mu\rho^\mu -i\chi (\eta_\mu\tilde\rho^\mu -
\lambda_2^\prime\tilde\chi)\right]
\end{eqnarray}
is the effective action in axial gauge with new gauge parameters $\lambda_1^\prime = 
(1+\gamma_2)\lambda_1 $ and $\lambda_2^\prime = (1+\gamma_2)\lambda_2 $.
Thus, the generalized BRST transformations with the finite parameter given in Eq. (\ref{fsym}) 
take
\begin{equation}
Z^L\left(=\int D\phi e^{iS_{eff}^L}\right)\stackrel{Generalized\ BRST}{-----\longrightarrow
} Z^A\left(=\int D\phi e^{iS_{eff}^A}\right).
\end{equation}
The effective theory of Abelian rank-2 antisymmetric field in axial gauge is convenient in
 many different situations. The generating functional in axial gauge with a suitable 
axis is same as the generating functional obtained by using Zwanziger's formulation for 
electric and magnetic charges \cite{zwan,deg}. Using the FFBRST transformations with the 
parameter 
given in Eq. (\ref{afin}) we have linked generating functionals in covariant and noncovariant 
gauges.
\subsection{Effective theory in Coulomb gauge}
The generating functional for the effective theories in Coulomb gauge can be obtained by using 
the generalized BRST transformations with a different parameter
\begin{eqnarray}
\Theta^\prime_b &=&\int d^4x\left[\gamma_1\tilde\rho_\nu (\partial_\mu B^{\mu\nu}-\partial_i 
B^{i\nu}-\partial^\nu\varphi )+\gamma_1\tilde\rho_i\partial^i\varphi +\gamma_2 
\lambda_1\tilde\rho_\nu\beta^\nu \right. \nonumber\\
&+&\left.\gamma_1\tilde\sigma (\partial_\mu\rho^\mu -\partial_i\rho^i )+\gamma_2 
\lambda_2\tilde\sigma\chi\right].\label{par}
\end{eqnarray} 
The effective action in Lorentz gauge, $S_{eff}^L$ as given in Eq. (\ref{act}) is invariant 
under generalized BRST transformations in Eq. (\ref{fsym}) corresponding to above mentioned
finite parameter. Now we consider the effect of this generalized BRST transformations on the 
generating functional in Lorentz gauge.

In Appendix B, it has been shown that the Jacobian for the path integral measure 
corresponding to this generalized BRST transformations can be replaced by $e^{iS_1^C}$, where 
$S_1^C$ is the local functional of fields calculated as
\begin{eqnarray}
S^C_1&=&\int d^4x \left [-\beta_\nu\partial_\mu B^{\mu\nu} +\beta_\nu\partial_i B^{i\nu} +\gamma_2 
\lambda_1\beta_\nu\beta^\nu -i\tilde\rho_\nu\partial_{\mu}(\partial^\mu\rho^\nu -
\partial^\nu\rho^\mu)\right.\nonumber\\
&+&\left. i\tilde\rho_\nu\partial_i(\partial^i\rho^\nu -\partial^\nu\rho^i )+i\gamma_2 
\lambda_2\chi\tilde\chi +i\tilde\chi\partial_\mu\rho^\mu -i
\tilde\chi\partial_i\rho^i\right.\nonumber\\
&+&\left. \tilde\sigma\partial_\mu\partial^\mu\sigma -\tilde\sigma\partial_i\partial^i\sigma - 
\partial_\mu\beta^\mu\varphi + \partial_i\beta^i\varphi +i\chi\partial_\mu\tilde\rho^\mu 
\right.\nonumber\\
&-&\left. i\chi\partial_i\tilde\rho^i\right],
\end{eqnarray}
 and this extra piece of the action can be added to the effective action in covariant gauge 
to lead a new effective action
\begin{eqnarray}
S^L_{eff} +S_1^C&=&\int d^4x\left[\frac{1}{12}F_{\mu\nu\rho}F^{\mu\nu\rho}+i
\tilde\rho_\nu\partial_i (\partial^i\rho^\nu -\partial^\nu\rho^i )-
\tilde\sigma\partial_i\partial^i\sigma +\beta_\nu(\partial_i B^{i\nu}\right.\nonumber\\ 
&+&\left. \lambda_1^\prime\beta^\nu) -\beta_i\partial^i\varphi 
- i\tilde\chi\partial_i\rho^i -i\chi (\partial_i\tilde\rho^i -\lambda_2^\prime\tilde\chi )
\right]\nonumber\\
&\equiv &S_{eff}^C,
\end{eqnarray}
which is an effective action in Coulomb gauge for Abelian rank 2 tensor field. Thus we can 
study the Abelian 2-form gauge theory in Coulomb gauge more rigorously through its connection 
with Lorentz gauge via finite BRST transformations.

\section{Generalized Anti-brst transformation in 2-form Gauge theories}
In this section, we consider the generalization of anti-BRST transformations following the 
similar method as discussed in section 4 and show it plays exactly similar role in 
connecting the generating functionals in different effective theories of Abelian rank-2 
antisymmetric tensor field.
\subsection{ Axial gauge theory using generalized anti-BRST}
For sake of convenience we recast the effective action in covariant gauge given in Eq. (\ref{act}) as
\begin{eqnarray}
S_{eff}^L&=&\int d^4x\left[\frac{1}{12}F_{\mu\nu\rho}F^{\mu\nu\rho}+i\partial_\mu\rho_\nu (
\partial^\mu\tilde\rho^\nu -\partial^\nu\tilde\rho^\mu )+
\partial_\mu\sigma\partial^\mu\tilde\sigma +\beta_\nu(\partial_\mu B^{\mu\nu} +
\lambda_1\beta^\nu -\partial^\nu\varphi)\right.\nonumber\\ 
&-&\left. i\chi\partial_\mu\tilde\rho^\mu -i\tilde\chi (\partial_\mu\rho^\mu +\lambda_2\chi )
\right],
\end{eqnarray}
which is invariant under following generalized anti-BRST transformations:
\begin{eqnarray}
\delta_{ab} B_{\mu\nu} &=& -(\partial_\mu\tilde\rho_\nu -\partial_\nu\tilde\rho_\mu)\Theta_{ab} [
\phi ] \nonumber\\
\delta_{ab}\tilde\rho_\mu &=& -i\partial_\mu\tilde\sigma \Theta_{ab} [\phi ],\ \ \ \ \ \ \ \ \ \ \ 
\ \ \ \ \delta_{ab}\tilde\sigma = 0 \nonumber\\
\delta_{ab}\rho_\mu &=&-i\beta_\mu \Theta_{ab} [\phi ],\ \ \ \ \ \ \ \ \ \ \ \ \ \ \ 
\delta_{ab}\beta_\mu = 0\nonumber\\
\delta_{ab}\sigma &=&-\chi\Theta_{ab} [\phi ],\ \ \ \ \ \ \ \ \ \ \ \ \ \ \ \ \ \ \ \delta_{ab}\chi 
=0\nonumber\\
\delta_{ab}\varphi &=& \tilde\chi\Theta_{ab} [\phi ],\ \ \ \ \ \ \ \ \ \ \ \ \ \ \ \ \ \ \ \ \ 
\delta_{ab}\tilde\chi =0, \label{antisym}
\end{eqnarray}
where $\Theta_{ab}$ is finite field dependent anti-BRST parameter.
To obtain the generating functional in axial gauge using finite field dependent anti-BRST (FF 
anti-BRST) transformations we choose,
\begin{eqnarray}
\Theta^\prime_{ab} &=&-\int d^4x\left[\gamma_1\rho_\nu (\partial_\mu B^{\mu\nu}-\eta_\mu B^{
\mu\nu}-\partial^\nu\varphi -\eta^\nu\varphi )+\gamma_2 \lambda_1\rho_\nu\beta^\nu \right. 
\nonumber\\
&-&\left.\gamma_1\sigma (\partial_\mu\tilde\rho^\mu -\eta_\mu\tilde\rho^\mu )+\gamma_2 
\lambda_2\sigma\tilde\chi\right]. \label{abfin}
\end{eqnarray}
This parameter is similar to $\Theta^\prime_b$ in Eq. (\ref{afin}) except the (anti)ghost and 
ghost 
of (anti)ghost fields are replaced by their antighost fields respectively.

Similar to FFBRST case, these finite field dependent anti-BRST transformations 
change the Jacobian in the path integral measure by a factor $e^{iS_1^A}$, 
where $S_1^A$ is a local 
functional of fields and given by
\begin{eqnarray}
S_1^A&=&\int d^4x [-\beta_\nu\partial_\mu B^{\mu\nu} +\beta_\nu\eta_\mu B^{\mu\nu} +\gamma_2
\lambda_1\beta_\nu\beta^\nu +i\rho_\nu\partial_{\mu}(\partial^\mu\tilde\rho^\nu -
\partial^\nu\tilde\rho^\mu)\nonumber\\
&-&i\rho_\nu\eta_{\mu}(\partial^\mu\tilde\rho^\nu -\partial^\nu\tilde\rho^\mu )+i\gamma_2 
\lambda_2\chi\tilde\chi +i\chi\partial_\mu\tilde\rho^\mu -i
\chi\eta_\mu\tilde\rho^\mu\nonumber\\
&+&\sigma\partial_\mu\partial^\mu\tilde\sigma -\sigma\eta_\mu\partial^\mu\tilde\sigma - 
\partial_\mu\beta^\mu\varphi + \eta_\mu\beta^\mu\varphi +i\tilde\chi\partial_\mu\rho^\mu 
\nonumber\\
&-&i\tilde\chi\eta_\mu\rho^\mu ].
\end{eqnarray}
It is easy to verify that
\begin{eqnarray}
S^L +S_1^A&=&\int d^4x\left[\frac{1}{12}F_{\mu\nu\rho}F^{\mu\nu\rho}-i\rho_\nu\eta_\mu (
\partial^\mu\tilde\rho^\nu -\partial^\nu\tilde\rho^\mu )-
\sigma\eta_\mu\partial^\mu\tilde\sigma +\beta_\nu(\eta_\mu B^{\mu\nu}\right.\nonumber\\ 
&+&\left. \lambda_1^\prime\beta^\nu +\eta^\nu\varphi) 
- i\tilde\chi\eta_\mu\rho^\mu -i\chi (\eta_\mu\tilde\rho^\mu -
\lambda_2^\prime\tilde\chi)\right]\nonumber\\
&\equiv &S^A,
\end{eqnarray}
which is the action in axial gauge in 2-form gauge theory.
Thus, the generalized anti-BRST transformations with the finite parameter given in Eq. (\ref{antisym}) 
take
\begin{equation}
Z^L\left(=\int D\phi e^{iS_{eff}^L}\right)\stackrel{Generalized\ anti-BRST}{-------
\longrightarrow
} Z^A\left(=\int D\phi e^{iS_{eff}^A}\right),
\end{equation}
which shows the FF-antiBRST transformations play a similar role as FFBRST transformations 
in 2-form gauge theroy.
\section{Field/Antifield formulation using FFBRST transformation in Abelian rank-2 Antisymmetric Tensor field theory}
In this section we construct field/antifield formulation for Abelian rank-2 antisymmetric
tensor field theory to show that techniques of FFBRST formulation can also be applied in
this modern approach of quantum field theory. For this purpose we express the 
generating functional in Eq. (\ref{zfun}) in field/antifield formulation by introducing 
antifield $\phi^\star $ corresponding to each field $\phi$ with opposite statistics as,
\begin{eqnarray}
Z^L &=&\int\left[dBd\rho d\tilde{\rho}d\sigma d\tilde{\sigma}d\varphi d\chi d\tilde{\chi}d
\beta\right]\exp\left[i\int d^4x\left\{\frac{1}{12}F_{\mu\nu\lambda}F^{\mu\nu\lambda}-B^{
\mu\nu\star}\left(\partial_\mu\rho_\nu-\partial_\nu\rho_\mu\right)\right.\right.\nonumber\\
&-&i\left.\left.\rho^{\mu\star}\partial_\mu\sigma +i{\tilde{\rho}}^{\nu\star}\beta_\nu-\tilde{
\sigma}^\star\tilde\chi-\varphi^\star\chi\right\}\right].
\end{eqnarray}
This can be written in compact form as
\begin{equation}
Z^L = \int D\phi \exp{\left [iW_{\Psi^L}(\phi,\phi^\star)\right]},
\end{equation}
where $W_{\Psi^L}(\phi,\phi^\star)$ is an extended action for 2-form gauge theory in Lorentz 
gauge corresponding the gauge fixed fermion $\Psi^L $  having grassman parity 1 and ghost 
number {-1}. The expression for
 $\Psi^L $ is
\begin{equation}
\Psi^L =-i\int d^4x \left[\tilde\rho_\nu \left(\partial_\mu B^{\mu\nu}+
\lambda_1\beta^\nu \right) +\tilde\sigma \partial_\mu\rho^\mu +\varphi 
\left(\partial_\mu\tilde{\rho}^
\mu-\lambda_2\tilde\chi \right)\right].
\end{equation}
 The generating functional $Z^L$ does not depend on the choice of gauge fixed fermion.
This extended quantum action, $W_{\Psi^L}(\phi,\phi^\star)$ satisfies certain rich mathematical
 relation called quantum master equation \cite{wei}, given by
\begin{equation}
\Delta e^{iW_\Psi[\phi, \phi^\star ]} =0  \ \mbox{ with }\ 
 \Delta\equiv \frac{\partial_r}{
\partial\phi}\frac{\partial_r}{\partial\phi^\star } (-1)^{\epsilon
+1}.
\label{mq}
\end{equation}
The antifields $\phi^\star$ corresponding to each field $\phi$ for this particular theory can 
be obtained from the gauge fixed fermion as
\begin{eqnarray}
B^{\mu\nu\star }&=&\frac{\delta \psi^L}{\delta B_{\mu\nu}}=i\partial^\mu\tilde\rho^\nu, \ \ \ \ \ \ \ \ \ \ \ \ \ 
\tilde\rho^{\nu\star}=\frac{\delta \psi^L}{\delta \tilde \rho_\nu}=-i(\partial_\mu B^{\mu\nu}+
\lambda_1\beta^\nu -\partial^\nu\varphi )\nonumber\\
\rho^{\mu\star }&=&\frac{\delta \psi^L}{\delta \rho_{\mu}}=i\partial^\mu\tilde\sigma, \ \ \ \ \ \  \ \ \ \ \ \ \ \ \ \
\tilde\sigma^\star =\frac{\delta \psi^L}{\delta\tilde\sigma}=-i\partial_\mu\rho^\mu\nonumber\\
\sigma^\star &=&\frac{\delta \psi^L}{\delta \sigma}=0, \ \ \ \ \ \ \ \ \ \ \ \ \ \ \ \ \ \ \ \ \beta^{\nu\star}=\frac{
\delta \psi^L}{\delta \beta_\nu}=-i\lambda_1\tilde\rho^\nu\nonumber\\
\varphi^\star &=&\frac{\delta \psi^L}{\delta\varphi}=-i(
\partial_\mu\tilde\rho^\mu-\lambda_2\tilde\chi ), \ \ \tilde\chi^\star =\frac{\delta \psi^L}{
\delta\tilde\chi}=i\lambda_2\varphi\nonumber\\
\chi^\star &=&\frac{\delta \psi^L}{\delta\chi}=0.
\end{eqnarray}
Now we apply the FFBRST transformation with the finite parameter given in Eq. (\ref{afin})
to this generating functional in Lorentz gauge. But the path integral measure is not invariant 
under such a finite transformation and give rise to a factor which can be written as 
$e^{iS_1}$, where the functional $S_1$ is calculated in Appendix A and also given in Eq. 
(\ref{s1}).
The transformed generating functional
\begin{eqnarray}
Z^\prime&=&\int D\phi \exp[i\{ W_{\Psi^L}+S_1\}],\nonumber\\
        &=&\int D\phi \exp[i W_{\Psi^A}]\equiv Z^A
\end{eqnarray}
The generating functional in axial gauge 
\begin{eqnarray}
Z^A &=&\int\left[dBd\rho d\tilde{\rho}d\sigma d\tilde{\sigma}d\varphi d\chi d\tilde{\chi}d
\beta\right ]\exp\left[i\int d^4x\left\{\frac{1}{12}F_{\mu\nu\lambda}F^{\mu\nu\lambda}-\bar B^{
\mu\nu\star}\left(\partial_\mu\rho_\nu-\partial_\nu\rho_\mu\right)\right.\right.\nonumber\\
&-&i\left.\left.\bar{\rho}^{\mu\star}\partial_\mu\sigma +i\bar{\tilde{\rho}}^{\nu\star}
\beta_\nu-\bar{\tilde{\sigma}}^\star\tilde\chi-\bar{\varphi}^\star\chi\right\}\right ].
\end{eqnarray}
The extended action $ W_{\Psi^A}$ for 2-form gauge theory in axial gauge also
satisfies  
the quantum master
equation (\ref{mq}).
The gauge fixed fermion for axial gauge   
\begin{equation}
\Psi^A =-i\int d^4x \left[\tilde\rho_\nu \left(\eta_\mu B^{\mu\nu}+
\lambda_1\beta^\nu \right) +\tilde\sigma \eta_\mu\rho^\mu +\varphi \left(\eta_\mu\tilde{\rho}^
\mu-\lambda_2\tilde\chi \right)\right].
\end{equation}
and corresponding antifields are
\begin{eqnarray}
 \bar B^{\mu\nu\star }&=&\frac{\delta \psi^A}{\delta B_{\mu\nu}}=-i\eta^\mu\tilde\rho^\nu, \ \ \ \ \ \ \ \ \ \ \ \ \  
\bar{\tilde\rho}^{\nu\star}=\frac{\delta \psi^A}{\delta \tilde \rho_\nu}=-i(\eta_\mu B^{\mu\nu
}+
\lambda_1^\prime\beta^\nu +\eta^\nu\varphi )\nonumber\\
\bar{\rho}^{\mu\star }&=&\frac{\delta \psi^A}{\delta \rho_{\mu}}=-i\eta^\mu\tilde\sigma, \ \ \ \ \ \  \ \ \ \ \ \ \ \ \ \ 
\bar{\tilde\sigma}^\star =\frac{\delta \psi^A}{\delta\tilde\sigma}=-i
\eta_\mu\rho^\mu\nonumber\\
\bar{\sigma}^\star &=&\frac{\delta \psi^A}{\delta \sigma}=0, \ \ \ \ \ \ \ \ \ \ \ \ \ \ \ \ \ \ \ \ \ \bar\beta^{
\nu\star}=\frac{\delta \psi^A}{\delta \beta_\nu}=-i\lambda_1^\prime\tilde\rho^\nu\nonumber\\
\bar\varphi^\star &=&\frac{\delta \psi^A}{\delta\varphi}=-i(
\eta_\mu\tilde\rho^\mu-\lambda_2^\prime\tilde\chi ), \ \ \ \bar{\tilde\chi}^\star =\frac{\delta 
\psi^A}{\delta\tilde\chi}=i\lambda_2^\prime\varphi\nonumber\\
\bar{\chi}^\star &=&\frac{\delta \psi^A}{\delta\chi}=0
\end{eqnarray}
Thus the FFBRST transformation with finite parameter given in Eq. (\ref{afin}) relates the 
different  solutions of quantum master equation
in field/antifield formulation.

\section{Concluding Remarks}
The usual BRST transformations have been generalized for the Abelian rank-2 antisymmetric 
tensor field theory by making the parameter finite and field dependent. Such generalized BRST 
transformations are nilpotent and leave the effective action invariant. However, being 
finite in nature such transformations do not leave the path integral measure invariant. We 
have shown that for certain choices of finite field dependent parameter, the Jacobian for the 
path integral of such generalized BRST transformations always can be written as $e^{iS_1}$, 
where $S_1$ is some local functional of fields and depends on the choice of the finite BRST 
parameter. $S_1$ can be added with $S_{eff}^L$ to produce the new effective action. Thus
the generating functional corresponding to one effective theory is then linked to the 
generating functional corresponding  to another effective theory through the generalized BRST 
transformations. In this present work we have shown that the generating functional 
corresponding to covariant gauge viz. Lorentz gauge is connected to the generating functional 
in noncovariant gauges viz. axial gauge and Coulomb gauge. Thus the generalized BRST 
transformations are helpful in the study of Abelian rank-2 antisymmetric tensor field theory 
in noncovariant gauges, which is very useful in certain situation \cite{deg}.
We further consider the generalization of anti-BRST transformations and show that even 
generalized anti-BRST transformations can connect generating functionals for different 
effective theories. The generalized BRST transformations can also be very useful 
in modern approach of quantum field theory, namely field/antifield 
formulation. With the help of an explicit example we have shown
 that the different solutions of the master 
equation are related through generalized BRST transformations in the field/antifield formulation
of Abelian 2-form antisymmetric tensor field theory.

\appendix

\section{FFBRST in Axial gauge}
Under the generalized BRST transformations with finite parameter given in Eq. (\ref{afin}), 
the path integral measure for the generating functional in Eq. (\ref{zfun}) transforms as 
\begin{equation}
\int D\phi^\prime =\int D\phi\; J(\kappa),
\end{equation}
$J(\kappa)$ can be replaced by $e^{iS_1^A}$ if the condition in Eq. (\ref{mcond}) satisfies.
We start with an ansatz for $S_1^A$ to connect the theory in Lorentz to axial gauge as
\begin{eqnarray}
S_1^A&=&\int d^4x \left[\xi_1\beta_\nu\partial_\mu B^{\mu\nu} +\xi_2\beta_\nu\eta_\mu B^{\mu\nu} +
\xi_3\beta_\nu\beta^\nu +i\xi_4\tilde\rho_\nu\partial_{\mu}(\partial^\mu\rho^\nu -
\partial^\nu\rho^\mu)\right.\nonumber\\
&+&\left. i\xi_5\tilde\rho_\nu\eta_{\mu}(\partial^\mu\rho^\nu -\partial^\nu\rho^\mu )+i
\xi_6\chi\tilde\chi +i\xi_7 \tilde\chi\partial_\mu\rho^\mu +i\xi_8 
\tilde\chi\eta_\mu\rho^\mu\right.\nonumber\\
&+&\left. \xi_9\tilde\sigma\partial_\mu\partial^\mu\sigma +\xi_{10}
\tilde\sigma\eta_\mu\partial^\mu\sigma +\xi_{11} \partial_\mu\beta^\mu\varphi +\xi_{12} 
\eta_\mu\beta^\mu\varphi\right.\nonumber\\
&+&i\left.\xi_{13}\chi\partial_\mu\tilde\rho^\mu +i\xi_{14}\chi\eta_\mu\tilde\rho^\mu \right] 
\label{ans1}
\end{eqnarray}
where $\xi_i(i=1, 2,...,14)$ are explicit $\kappa$ dependent parameters to be determined by 
using Eq. (\ref{mcond}).
The infinitesimal change in Jacobian, using Eq. (\ref{jac}) with finite parameter in Eq. (\ref
{afin}), is calculated as
\begin{eqnarray}
\frac{1}{J}\frac{dJ}{d\kappa}=-\int &d^4x&\left[-i\gamma_1\beta_\nu (\partial_\mu B^{\mu\nu}-
\eta_\mu B^{\mu\nu})+\gamma_1\tilde\rho_\nu\partial_\mu (\partial^\mu\rho^\nu -
\partial^\nu\rho^\mu )\right.\nonumber\\
&-&\left.\gamma_1\tilde\rho_\nu\eta_\mu (\partial^\mu\rho^\nu -\partial^\nu\rho^\mu )-i
\gamma_2 \lambda_1\beta_\nu\beta^\nu -\gamma_1\tilde\chi\partial_\mu\rho^\mu \right.\nonumber\\
&+&\left.\gamma_1\tilde\chi\eta_\mu\rho^\mu +\gamma_2 \lambda_2\chi\tilde\chi +i
\gamma_1\tilde\sigma\partial_\mu\partial^\mu\sigma -i
\gamma_1\tilde\sigma\eta_\mu\partial^\mu\sigma\right.\nonumber\\
&+&\left. i\gamma_1\beta_\mu(\partial^\mu\varphi +\eta^\mu\varphi) -
\gamma_1\chi\partial_\mu\tilde\rho^\mu +\gamma_1\chi\eta_\mu\tilde\rho^\mu\right].
\end{eqnarray}

The condition (\ref{mcond}) will be satisfied iff
\begin{eqnarray}
\int &d^4x&e^{i(S_{eff}+S_1^A)}\left [i\beta_\nu\partial_\mu B^{\mu\nu}(\xi_1^\prime -\gamma_1
 )+i\beta_\nu\eta_\mu B^{\mu\nu}(\xi_2^\prime +\gamma_1 )+i\beta_\nu\beta^\nu (\xi_3^\prime -
\gamma_2 \lambda_1)\right.\nonumber\\
&-&\left.\tilde\rho_\nu\partial_\mu (\partial^\mu\rho^\nu-\partial^\nu\rho^\mu )(\xi_4^\prime 
-\gamma_1 )-\tilde\rho_\nu\eta_\mu (\partial^\mu\rho^\nu-\partial^\nu\rho^\mu )(\xi_5^\prime +
\gamma_1 )-\chi\tilde\chi (\xi_6^\prime -\gamma_2 \lambda_2)\right.\nonumber\\
&-&\left.\tilde\chi\partial_\mu\rho^\mu (\xi_7^\prime +\gamma_1 )-\tilde\chi\eta_\mu\rho^\mu (
\xi_8^\prime -\gamma_1 ) +i\tilde\sigma\partial_\mu\partial^\mu\sigma (\xi_9^\prime +\gamma_1 
)\right.\nonumber\\
&+&\left.\tilde\sigma\eta_\mu\partial^\mu\sigma (\xi_{10}^\prime -\gamma_1 )+i
\partial_\mu\beta^\mu\varphi (\xi_{11}^\prime -\gamma_1 )+i\eta_\mu\beta^\mu\varphi (\xi_{12}^
\prime +\gamma_1 )\right.\nonumber\\
&-&\left.\chi\partial_\mu\tilde\rho^\mu (\xi_{13}^\prime +\gamma_1 )-
\chi\eta_\mu\tilde\rho^\mu (\xi_{14}^\prime -\gamma_1 )+i\partial_\mu (
\partial^\mu\rho^\nu-\partial^\nu\rho^\mu )\Theta^\prime_b [
\beta_\nu(\xi_4-\xi_1)]\right.\nonumber\\
&+&\left.i\eta_\mu (\partial^\mu\rho^\nu-\partial^\nu\rho^\mu )\Theta^\prime_b [
\beta_\nu(\xi_5-\xi_2)] -i\partial_\mu\partial^\mu\sigma\Theta^\prime_b [\tilde\chi(\xi_7 -
\xi_9)]\right.\nonumber\\
&-&\left.i\eta_\mu\partial^\mu\sigma\Theta^\prime_b [\tilde\chi (\xi_8 -\xi_{10})]-i
\partial^\mu\chi\Theta^\prime_b [\beta_\mu (\xi_{11}+\xi_{13})]\right.\nonumber\\
&-&\left. i\eta^\mu\chi\Theta^\prime_b [\beta_\mu (\xi_{12}+\xi_{14})]\right] =0\label{mcond1}
\end{eqnarray}

The contribution of antighost and ghost of antighost can possibly vanish by using the 
equations of motion of the $\tilde\rho_\mu$ and $\tilde\sigma$. It will happen if the ratio of 
the coefficient of terms in the above equation and the similar terms in $S^L_{eff} +S_1^A$ is 
identical \cite{sdj4}.
This requires that
\begin{eqnarray}
\frac{\xi_4^\prime -\gamma_1}{\xi_4 +1}&=&\frac{\xi_5^\prime +\gamma_1}{\xi_5}\nonumber\\
\frac{\xi_9^\prime +\gamma_1}{\xi_9 -1}&=&\frac{\xi_{10}^\prime -\gamma_1}{\xi_{10}}
\label{req1}
\end{eqnarray}
The $\Theta^\prime$ dependent terms can be converted into local terms by the equation of 
motion of different fields. This can only work if the following conditions are satisfied
\begin{eqnarray}
\frac{\xi_4-\xi_1}{\xi_4+1}&=&\frac{\xi_5 -\xi_2}{\xi_5}\nonumber\\
\frac{\xi_7-\xi_9}{\xi_9 -1}&=&\frac{\xi_8 -\xi_{10}}{\xi_{10}}\nonumber\\
\frac{\xi_{11}+\xi_{13}}{\xi_{13} -1}&=&\frac{\xi_{12} +\xi_{14}}{\xi_{14}}
\label{req2}
\end{eqnarray}
Further by comparing the coefficients of different terms $ 
i\beta_\nu\partial_\mu B^{\mu\nu}$, $i\beta_\nu\eta_\mu B^{\mu\nu}$, $i\beta_\nu\beta^\nu$, $
\tilde\rho_\nu\partial_\mu (\partial^\mu\rho^\nu-\partial^\nu\rho^\mu )$, $
\tilde\rho_\nu\eta_\mu (\partial^\mu\rho^\nu-\partial^\nu\rho^\mu )$, $
\chi\tilde\chi$, $\tilde\chi\partial_\mu\rho^\mu$, $\tilde\chi\eta_\mu\rho^\mu$, $ 
i\tilde\sigma\partial_\mu\partial^\mu\sigma$, $i\tilde\sigma\eta_\mu\partial^\mu\sigma$,
 $i\partial_\mu\beta^\mu\varphi$, $i\eta_\mu\beta^\mu\varphi$, $\chi\partial_\mu\tilde\rho^\mu$
and $\chi\eta_\mu\tilde\rho^\mu$ in both sides of Eq. (\ref{mcond1}), we obtain the following conditions
\begin{eqnarray}
&&\xi_1^\prime  -\gamma_1 +\gamma_1 (\xi_4 -\xi_1 )+\gamma_1 (\xi_5 -\xi_2 )=0\nonumber\\
&&\xi_2^\prime +\gamma_1 -\gamma_1 (\xi_4 -\xi_1 )-\gamma_1 (\xi_5 -\xi_2 )=0\nonumber\\
&&\xi_3^\prime -\gamma_2 \lambda_1 +\gamma_2 \lambda_1(\xi_4 -\xi_1 )+\gamma_2 \lambda_1 (
\xi_5 -\xi_2 )=0\nonumber\\
&&\xi_4^\prime -\gamma_1 =0\nonumber\\
&&\xi_5^\prime +\gamma_1 =0\nonumber\\
&&\xi_6^\prime -\gamma_2 \lambda_2 -\gamma_2 \lambda_2(\xi_7 -\xi_9 )-\gamma_2 \lambda_2 (
\xi_8 -\xi_{10} )=0\nonumber\\
&&\xi_7^\prime +\gamma_1 -\gamma_1 (\xi_7 -\xi_9 )-\gamma_1 (\xi_8 -\xi_{10})=0\nonumber\\
&&\xi_8^\prime -\gamma_1 +\gamma_1 (\xi_7 -\xi_9 )+\gamma_1 (\xi_8 -\xi_{10})=0\nonumber\\
&&\xi_9^\prime +\gamma_1 =0\nonumber\\
&&\xi_{10}^\prime -\gamma_1 =0\nonumber\\
&&\xi_{11}^\prime -\gamma_1 -\gamma_1 (\xi_{11} +\xi_{13} )-\gamma_1 (\xi_{12} +\xi_{14})=0
\nonumber\\
&&\xi_{12}^\prime +\gamma_1 +\gamma_1 (\xi_{11} +\xi_{13} )+\gamma_1 (\xi_{12} +\xi_{14})=0
\nonumber\\
&&\xi_{13}^\prime +\gamma_1 =0\nonumber\\
&&\xi_{14}^\prime -\gamma_1 =0.
\end{eqnarray}
A particular solution of the above differential equations subjected to the conditions in
Eqs. (\ref{req1}) and (\ref{req2}) with initial condition $\xi_i(\kappa=0)=0$ 
can be written as 
\begin{eqnarray}
\xi_1 &=&-\kappa,\ \ \ \ \ \ \xi_2 =\kappa\nonumber\\
\xi_3 &=&\gamma_2 \lambda_1\kappa,\ \ \ \xi_4 =-\kappa\nonumber\\
\xi_5 &=&\kappa,\ \ \ \ \ \ \ \xi_6 =\gamma_2 \lambda_2\kappa\nonumber\\
\xi_7 &=&\kappa,\ \ \ \ \ \ \ \xi_8 =-\kappa\nonumber\\
\xi_9 &=&\kappa,\ \ \ \ \ \ \ \xi_{10} =-\kappa\nonumber\\
\xi_{11} &=&-\kappa,\ \ \ \ \ \xi_{12} =\kappa\nonumber\\
\xi_{13} &=&\kappa,\ \ \ \ \ \xi_{14} =-\kappa, \label{soln}
\end{eqnarray}
where we have chosen the arbitrary parameter $\gamma_1 =-1$.
Putting the values of $\xi_i$ in Eq. (\ref{ans1}) we obtain $S_1^A$ at $\kappa =1$ as
\begin{eqnarray}
S_1^A&=&\int d^4x \left[-\beta_\nu\partial_\mu B^{\mu\nu} +\beta_\nu\eta_\mu B^{\mu\nu} +\gamma_2
\lambda_1\beta_\nu\beta^\nu -i\tilde\rho_\nu\partial_{\mu}(\partial^\mu\rho^\nu -
\partial^\nu\rho^\mu)\right.\nonumber\\
&+&\left. i\tilde\rho_\nu\eta_{\mu}(\partial^\mu\rho^\nu -\partial^\nu\rho^\mu )+i\gamma_2 
\lambda_2\chi\tilde\chi +i\tilde\chi\partial_\mu\rho^\mu -i
\tilde\chi\eta_\mu\rho^\mu\right.\nonumber\\
&+&\left. \tilde\sigma\partial_\mu\partial^\mu\sigma -\tilde\sigma\eta_\mu\partial^\mu\sigma - 
\partial_\mu\beta^\mu\varphi + \eta_\mu\beta^\mu\varphi +i\chi\partial_\mu\tilde\rho^\mu 
\right.\nonumber\\
&-&\left. i\chi\eta_\mu\tilde\rho^\mu \right].
\end{eqnarray}

\section{FFBRST in Coulomb gauge}
For the finite parameter given in Eq. (\ref{par})  we make following ansatz for $S_1^C$ 

\begin{eqnarray}
S_1^C&=&\int d^4x \left[\xi_1\beta_\nu\partial_\mu B^{\mu\nu} +\xi_2\beta_\nu\partial_i B^{i\nu} +
\xi_3\beta_\nu\beta^\nu +i\xi_4\tilde\rho_\nu\partial_{\mu}(\partial^\mu\rho^\nu -
\partial^\nu\rho^\mu)\right.\nonumber\\
&+&\left. i\xi_5\tilde\rho_\nu\partial_i(\partial^i\rho^\nu -\partial^\nu\rho^i )+i
\xi_6\chi\tilde\chi +i\xi_7 \tilde\chi\partial_\mu\rho^\mu +i\xi_8 
\tilde\chi\partial_i\rho^i\right.\nonumber\\
&+&\left. \xi_9\tilde\sigma\partial_\mu\partial^\mu\sigma +\xi_{10}
\tilde\sigma\partial_i\partial^i\sigma +\xi_{11} \partial_\mu\beta^\mu\varphi +\xi_{12} 
\partial_i\beta^i\varphi\right.\nonumber\\
&+&\left. i\xi_{13}\chi\partial_\mu\tilde\rho^\mu +i\xi_{14}\chi\partial_i\tilde\rho^i \right].
\end{eqnarray}

The infinitesimal change in Jacobian for Coulomb gauge is calculated as
\begin{eqnarray}
\frac{1}{J}\frac{dJ}{d\kappa}=-\int &d^4x&\left[-i\gamma_1\beta_\nu (\partial_\mu B^{\mu\nu}-
\partial_i B^{i\nu})+\gamma_1\tilde\rho_\nu\partial_\mu (\partial^\mu\rho^\nu -
\partial^\nu\rho^\mu )\right.\nonumber\\
&-&\left.\gamma_1\tilde\rho_\nu\partial_i (\partial^i\rho^\nu -\partial^\nu\rho^i )-i\gamma_2
\lambda_1\beta_\nu\beta^\nu -\gamma_1\tilde\chi\partial_\mu\rho^\mu \right.\nonumber\\
&+&\left.\gamma_1\tilde\chi\partial_i\rho^i +\gamma_2 \lambda_2\chi\tilde\chi +i
\gamma_1\tilde\sigma\partial_\mu\partial^\mu\sigma -i
\gamma_1\tilde\sigma\partial_i\partial^i\sigma\right.\nonumber\\
&+&\left. i\gamma_1\beta_\mu\partial^\mu\varphi -i\gamma_1\beta_i\partial^i\varphi -
\gamma_1\chi\partial_\mu\tilde\rho^\mu +\gamma_1\chi\partial_i\tilde\rho^i\right].
\end{eqnarray}
The condition will be satisfied iff
\begin{eqnarray}
\int &d^4x&e^{i(S_{eff}+S_1^C)}\left [i\beta_\nu\partial_\mu B^{\mu\nu}(\xi_1^\prime -\gamma_1
 )+i\beta_\nu\partial_i B^{i\nu}(\xi_2^\prime +\gamma_1 )+i\beta_\nu\beta^\nu (\xi_3^\prime -
\gamma_2 \lambda_1)\right.\nonumber\\
&-&\left.\tilde\rho_\nu\partial_\mu (\partial^\mu\rho^\nu-\partial^\nu\rho^\mu )(\xi_4^\prime 
-\gamma_1 )-\tilde\rho_\nu\partial_i (\partial^i\rho^\nu -\partial^\nu\rho^i )(\xi_5^\prime +
\gamma_1 )-\chi\tilde\chi (\xi_6^\prime -\gamma_2 \lambda_2)\right.\nonumber\\
&-&\left.\tilde\chi\partial_\mu\rho^\mu (\xi_7^\prime +\gamma_1 )-\tilde\chi\partial_i\rho^i (
\xi_8^\prime -\gamma_1 )+i\tilde\sigma\partial_\mu\partial^\mu\sigma (\xi_9^\prime +\gamma_1 )
\right.\nonumber\\
&+&\left.\tilde\sigma\partial_i\partial^i\sigma (\xi_{10}^\prime -\gamma_1 )+i
\partial_\mu\beta^\mu\varphi (\xi_{11}^\prime -\gamma_1 )+i\partial_i\beta^i\varphi (\xi_{12}^
\prime +\gamma_1 )\right.\nonumber\\
&-&\left.\chi\partial_\mu\tilde\rho^\mu (\xi_{13}^\prime +\gamma_1 )-
\chi\partial_i\tilde\rho^i (\xi_{14}^\prime -\gamma_1 )+i\partial_\mu (
\partial^\mu\rho^\nu-\partial^\nu\rho^\mu )\Theta^\prime_b [
\beta_\nu(\xi_4-\xi_1)]\right.\nonumber\\
&+&\left.i\partial_i(\partial^i\rho^\nu-\partial^\nu\rho^i )\Theta^\prime_b [
\beta_\nu(\xi_5-\xi_2)] -i\partial_\mu\partial^\mu\sigma\Theta^\prime_b [\tilde\chi(\xi_7 -
\xi_9)]\right.\nonumber\\
&-&\left.i\partial_i\partial^i\sigma\Theta^\prime_b [\tilde\chi (\xi_8 -\xi_{10})]-
\partial^\mu\chi\Theta^\prime_b [\beta_\mu (\xi_{11}+\xi_{13})]\right.\nonumber\\
&-&\left. i\partial^i\chi\Theta^\prime_b [\beta_i (\xi_{12}+\xi_{14})]\right] =0.
\end{eqnarray}
Following the method exactly similar to Appendix A, we obtain the solution for the parameters $\xi_i$ which is exactly same as in Eq. (\ref{soln}). 

Thus we obtain $S_1^C$ at $\kappa =1$ as
\begin{eqnarray}
S^C_1&=&\int d^4x \left[-\beta_\nu\partial_\mu B^{\mu\nu} +\beta_\nu\partial_i B^{i\nu} +\gamma_2
\lambda_1\beta_\nu\beta^\nu -i\tilde\rho_\nu\partial_{\mu}(\partial^\mu\rho^\nu -
\partial^\nu\rho^\mu)\right.\nonumber\\
&+&\left. i\tilde\rho_\nu\partial_i(\partial^i\rho^\nu -\partial^\nu\rho^i )+i\gamma_2 
\lambda_2\chi\tilde\chi +i\tilde\chi\partial_\mu\rho^\mu -i
\tilde\chi\partial_i\rho^i\right.\nonumber\\
&+&\left. \tilde\sigma\partial_\mu\partial^\mu\sigma -\tilde\sigma\partial_i\partial^i\sigma - 
\partial_\mu\beta^\mu\varphi + \partial_i\beta^i\varphi +i\chi\partial_\mu\tilde\rho^\mu 
\right.\nonumber\\
&-&\left. i\chi\partial_i\tilde\rho^i \right].
\end{eqnarray}

\section*{Acknowledgments}

We thankfully acknowledge the financial support from the Department of Science and Technology 
(DST), Government of India, under the SERC project sanction grant No. SR/S2/HEP-29/2007.

\end{document}